# Validation of Modulation Transfer Functions and Noise Power Spectra from Natural Scenes

Edward W. S. Fry*[1], Sophie Triantaphillidou[1], Robin B. Jenkin[2], John, R. Jarvis1, Ralph E. Jacobson[1]

[1]Computational Vision and Imaging Technology Research Group, University of Westminster, London, UK
[2]NVIDIA Corporation, Santa Clara, CA, USA
*Corresponding author (Email: e.fry@my.westminster.ac.uk)

## ABSTRACT

The Modulation Transfer Function (MTF) and the Noise Power Spectrum (NPS) characterize imaging system sharpness/resolution and noise, respectively. Both measures are based on linear system theory but are applied routinely to systems employing non-linear, content-aware image processing. For such systems, MTFs/NPSs are derived inaccurately from traditional test charts containing edges, sinusoids, noise or uniform tone signals, which are unrepresentative of natural scene signals. The dead leaves test chart delivers improved measurements, but still has limitations when describing the performance of scene-dependent systems. In this paper, we validate several novel scene-and-process-dependent MTF (SPD-MTF) and NPS (SPD-NPS) measures that characterize, either: i) system performance concerning one scene, or ii) average real-world performance concerning many scenes, or iii) the level of system scene-dependency. We also derive novel SPD-NPS and SPD-MTF measures using the dead leaves chart. We demonstrate that all the proposed measures are robust and preferable for scene-dependent systems than current measures.

Keywords: Modulation Transfer Function (MTF), Noise Power Spectrum (NPS), Scene-Dependency, Non-Linearity, Resolution, Sharpness, Noise, Scene and Process Dependent MTF (SPD-MTF), Scene and Process Dependent NPS (SPD-NPS)

## Introduction

This paper is concerned with the characterization of the spatial imaging performance of capturing systems, as implemented in their design, engineering, and image quality modelling. The *Modulation Transfer Function* (MTF) and *Noise Power Spectrum* (NPS) are commonly employed measures for such purposes, and measure system signal transfer (relating to sharpness and resolution attributes [1]) and noise, respectively. The MTF and the NPS are both defined in the next section. They are core input parameters in both univariate [2] and multivariate image quality metrics (IQM) [1], [3], [4]. They are derived traditionally by capturing test charts that provide well-characterized, relevant input signals for subsequent comparison with the output signal of the capturing system. All measurement methods aim to describe the "general" performance of the system in real use cases (i.e. its average performance when capturing natural scenes).

The MTF and NPS are based on the Fourier theory of image formation [5]. They rely on linear system theory, and should strictly only be measured from linear, spatially invariant and homogenous systems [5]. When they are derived from such systems, in theory, they fully specify the system, as shown in Figure 1a. This means that measurements derived from any test chart, $s_n$, describe average real-world system performance accurately, $F(s)$. In practice, however, such linear, spatially invariant and homogenous systems do not behave to these constraints, due to measurement error.

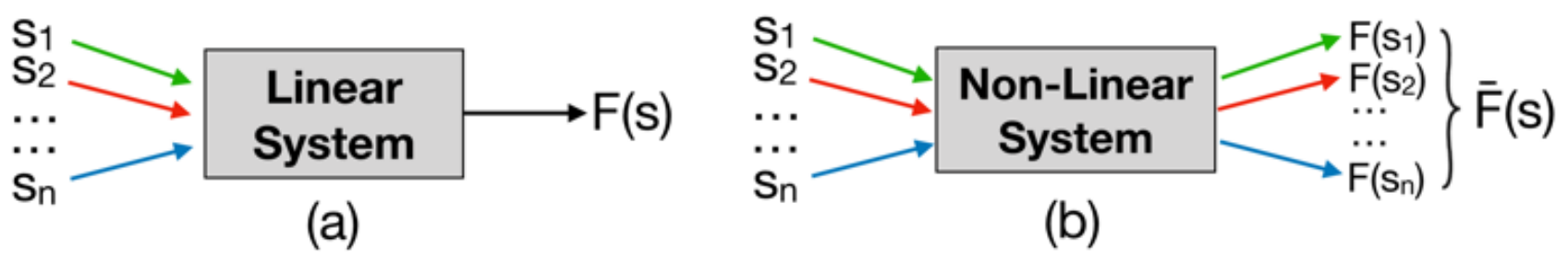

*Figure 1. Illustration of theoretical properties of the MTF or NPS ($F(s)$) for signals ($s$) present in a range of $n$ natural scenes (or test charts): (a) linear system, (b) non-linear system. $\bar{F}(s)$ is the mean of $F(s_1)$ to $F(s_n)$.*

More importantly, measurements from systems that apply non-linear, content-aware image processing are dependent on the content of the input signal – important processes to consider are noise reduction and sharpening. These "scene-dependent" systems are thus significantly more difficult to characterize. Such systems are increasingly common and include camera-phones as well as camera systems used by autonomous vehicles. Figure 1b defines their average real-world performance, $\bar{F}(s)$, as the mean of all MTFs/NPSs from the infinite number of potential natural scenes that the system may capture, $s_n$.

Classical measuring techniques used to derive the MTF employ test charts that consist of sinusoids [5]–[9], edges [5], [9], [10] and random noise signals [11]. These signals provide information suitable for characterizing the real-world performance of linear systems, but they are not representative of natural scene signal content. As a result, for non-linear capturing systems, the use of different charts often yields significantly different MTFs [12], which are unrepresentative of the performance of the system in real capturing scenarios. Likewise, NPSs derived traditionally from uniform luminance patches are unrepresentative of the average real-world noise performance of such systems, for the same reason [13].

The more recently introduced *dead leaves* test chart provides a more appropriate signal for the measurement of signal transfer [14]–[16] and noise [13] in non-linear systems and represents a step toward measuring performance using natural scenes. It simulates natural scene textures using a stochastic model [17], i.e. a series of circles of random intensity and size are overlaid, reproducing occlusion phenomena and varying contrast levels. It models the inverse power function of the "average" natural scene, among other natural scene statistics (NSS) [14]. It is designed with the aim of triggering non-linear image processing algorithms at comparable levels to when they are triggered when capturing natural scenes.

To-date, the NPS has not been measured directly using the dead leaves chart, although noise has been measured by comparing MTFs from different dead leaves implementations [13]. Such noise measures have not been used as input parameters in the *direct dead leaves MTF implementation* [15], Eq. 3, and should, in theory, be more appropriate than the uniform patch NPSs currently employed.

No measurement from any test chart can represent the MTF/NPS of non-linear systems with respect to a given input scene. Such an MTF/NPS is defined as $F(s_1)$ in Figure 1b, if $s_1$ is the signal of the scene. Branca et al. have made a first attempt to derive MTFs from natural scenes with some success [18]. But no prior art has combined MTFs/NPSs derived from many scenes, $F(s_n)$, to yield measurements of average real-world system performance, $\bar{F}(s)$, that account for scene-dependency. The field also lacks measures for the degree of scene-dependent variation in system performance. This can be defined as the spread of all MTFs/NPSs, $F(s_n)$, produced by the infinite number of potential natural scenes, $s_n$, in Figure 1b.

This paper proposes *scene-and-process-dependent MTF (SPD-MTF) and NPS (SPD-NPS)* measurement frameworks that account for the effect the input image content has upon any employed non-linear content-aware image signal processing. There are four SPD-MTF measures, and four SPD-NPS measures, which use different input signals. Each measure tests one of the following hypotheses:

1) The NPS can be derived from dead leaves signals and is more appropriate than the uniform patch NPS.
2) The accuracy of direct dead leaves MTFs [15], Eq. 3, improves if they compensate for noise in the system using #1.
3) The MTF/NPS can be derived from any given natural input scene while accounting for system scene-dependency.
4) The average real-world performance of the system can be characterized, while accounting for its scene-dependency, as the mean of #3 over a large and suitably varied image set.
5) The level of system scene-dependency can be measured as the standard deviation of #3 over the same image set.

All measures are validated by evaluating measurements from simulated linear and non-linear image capture pipelines. A question remains on whether such scene-and-process-dependent performance measures improve the accuracy of spatial IQMs that use them as input parameters, e.g. Barten's [4] *Square Root Integral with Noise (SQRIn)* and the IEEE P1858 *Camera Phone Image Quality*

*(CPIQ) metric* [1]. This question has been addressed in reference [19] by the same authors, indicating the measures' suitability and potential.

The following sections of this paper define the MTF and NPS and describe the current measurement methods and their limitations briefly. The SPD-MTF and SPD-NPS measures are then presented, including their sources of error, advantages and disadvantages. The test image dataset and simulated pipelines are described later. All proposed measures are then validated; sources of error and scene-dependent pipeline behavior are presented. Finally, we draw conclusions on the validity of our hypotheses, the proposed measures and their broader application and merits.

## Background on MTF and NPS Measures

### *The Noise Power Spectrum (NPS)*

The NPS characterizes the power of noise introduced by the system, with respect to spatial frequency, $u$. Equation 1 defines the one-dimensional (1D) NPS, $NPS(u)$, as the squared modulus of the Fourier transform of fluctuations from the mean signal of a 1D noise trace, $\Delta D(x)$, divided by the integration range, $x$ [5].

$$NPS(u) = \frac{1}{x}\left|\int_0^x \Delta D(x) e^{-2\pi i u x}\, dx\right|^2 \qquad (1)$$

The 1D NPS can be obtained from digital systems as the radial average of a two-dimensional (2D) NPS computed using the discrete Fourier transform (DFT), shown in Eq. 2.1, where $\mathrm{u}$ and $\mathrm{v}$ are spatial frequencies. $\mathrm{H}(\mathrm{x},\mathrm{y})$ is a noise image of dimensions $\mathrm{M}$ by $\mathrm{N}$, as expressed by Eq. 2.2, where $\mathrm{g}(\mathrm{x},\mathrm{y})$ is the output image intensity, and $\bar{\mathrm{g}}(\mathrm{x},\mathrm{y})$ is the expected (or mean signal) value.

$$NPS(u,v) = \left| \sum_{x=-\frac{M}{2}+1}^{M/2} \sum_{y=-\frac{N}{2}+1}^{N/2} H(x,y) e^{-2\pi i(ux+vy)} \right|^2 \tag{2.1}$$

$$where \quad H(x,y) = g(x,y) - \bar{g}(x,y) \tag{2.2}$$

When characterizing image capture systems, the NPS is normally derived from captured uniform luminance patches. This renders $\bar{g}(x,y)$ approximately constant at all coordinates if lens shading correction is applied, or if patches are captured in areas where lens shading is minimal. However, the method assumes that that noise present in captured uniform patches is representative of the noise in captured pictorial images. This assumption fails for all capture systems since photon noise magnitude is a function of the intensity of the scene. It fails more noticeably, however, for systems that apply non-linear content-aware denoising and sharpening.

Linear denoising algorithms 'average-out' noise at the expense of image detail, texture and edge contrast. Non-linear content-aware algorithms adjust their denoising intensity to preserve valuable signal content. Spatial domain examples apply thresholding in the presence of luminance gradients [20], [21]. Other examples operate adaptively in alternative domains [22]–[24], or implement machine learning [25]–[29] to denoise certain features more than others. Structural signals impede the local removal of noise by such algorithms, rendering system noise scene-dependent [3], [13], as shown in Figure 2. Uniform patches provide ideal input conditions for these algorithms. Thus, when captured, the former are generally less noisy than real scenes, Figure 2, and NPSs derived from them underestimate the level of system noise present in real capturing scenarios. Non-linear content-aware sharpening algorithms cause further scene-dependency. They amplify the contrast of edges, detail and high frequencies selectively – as described in the next sub-section – increasing the noise in these regions more than in others.

The SPD-NPS measurement framework proposed in this paper aims to account for the above scene-dependent behavior.

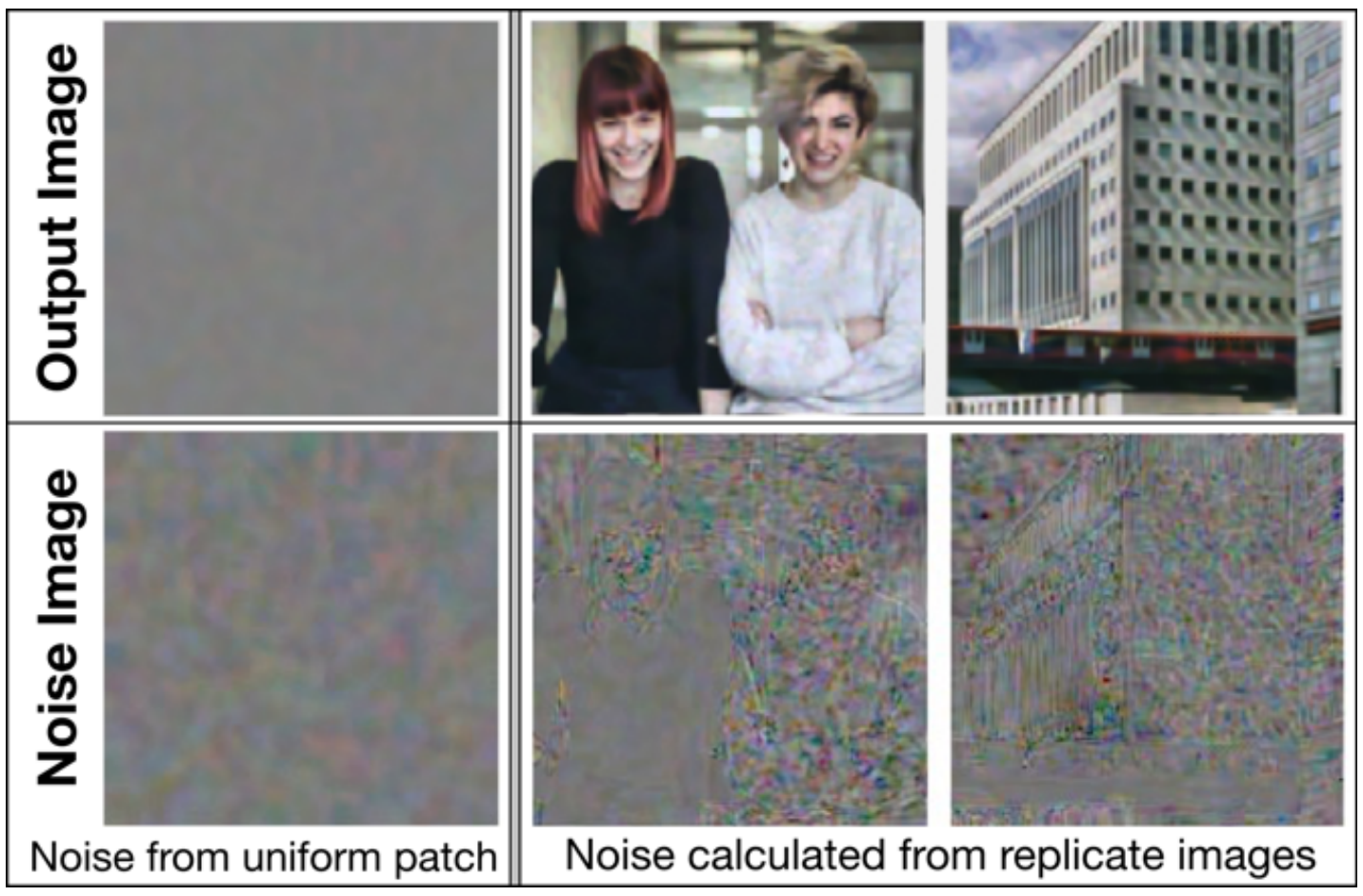

*Figure 2. Noise images generated from a non-linear camera simulation pipeline at SNR 10* [3], *for the following: uniform patch (left), Branca et al.* [18] *'People' image (center) and 'Architecture' image (right). Noise image contrast was increased to emphasize scene-dependent noise variation.*

***The Modulation Transfer Function (MTF)***

The MTF, and related *Spatial Frequency Response* (SFR), characterize the reproduction of modulation (image contrast) versus spatial frequency. The 1D MTF is defined as the modulus of the Fourier transform of the line spread function (LSF) of the system, which is the 1D integral of the system's point spread function (PSF) [5]. Since denoising removes edges, detail and texture, the adaptive characteristics of non-linear content-aware denoising algorithms introduce scene-dependency to system signal-transfer. Non-linear content-aware sharpening introduces further scene-dependency. For example, adaptive unsharp-masks (USM) [30]–[33] reduce contrast amplification in areas of a low signal gradient to avoid boosting noise. Other filters perform content-aware operations in various domains [34]–[37] that may employ guidance images [36], or multi-scale contrast manipulation [37].

Capture system MTFs are measured traditionally by the methods described below. Firstly, the LSF can be derived from edge signals [5], [10]. The ISO 12233:2017 [9] slanted-edge method is most common; it has been shown to describe well the subjective sharpness of non-linear systems [12]. However, signals from edge test charts are generally denoised less and sharpened more than natural scene signals by the non-linear, content aware de-noising and sharpening algorithms incorporated in many contemporary capturing systems. The MTF can also be measured from sinusoidal frequencies of interest [5]–[7], or other sine-wave charts such as the ISO 12233:2017 Siemens Star [9]. Sinusoidal signals respond less to sharpening and describe well the limiting resolution of non-linear systems [12]. The MTF may also be derived as the first order Wiener kernel transform of white noise signals [11].

MTFs derived from edges, sinusoidal signals and noise by the above methods each have different sources and levels of measurement error. This includes variation error resulting from: i) inaccurate specification of the input (target) signal, ii) inaccurate measurement of the output signal, and iii) errors resulting from processing when computing the MTF (e.g. DFT computation). Consequently, although a unique MTF exists for linear systems in theory, Figure 1a, in practice each target produces a different result.

Further, the differences obtained in MTF results by implementing measurement techniques from different test charts is amplified when it comes to the characterization of cameras that apply non-linear content-aware sharpening and denoising processes, since such processes react differently to input edges, sinusoids and random noise. The signal transfer of systems that apply such algorithms is, therefore, target-dependent (i.e. dependent upon the signal of the test chart) – which we more generally refer to in this paper as scene-dependency.

Since edges, sinusoidal signals and noise have little relation to the average pictorial scene, their interaction with non-linear content-aware processing means that the derived MTFs consistently over-estimate or under-estimate the average real-world system performance, shown as $\bar{F}(s)$ in Figure 1b, i.e. they are biased. MTFs derived from such signals also fail to describe texture loss [12], [14]–[16] that is a primary driver of perceived capture system image quality [38].

The above facts lead us to conclude that in order to characterize real-world camera performance MTFs should be measured from non-linear systems using targets that reflect the signal properties of pictorial scenes, or even better pictorial scenes themselves, provided that this does not significantly increase levels of measurement error.

MTFs are nowadays commonly measured from non-linear systems using the dead leaves chart [14]–[16] that relates more closely to natural scene signals than edges, sinusoidal signals and noise, and has been demonstrated to characterize texture loss effectively.

The direct dead leaves implementation for measuring the MTF [15] is widely adopted, and core to the IEEE P1858 texture acutance metric and multivariate IQM [1]. It is defined using Eq. 3. It compensates for the system's noise power, $NPS_{Output}(u)$, using the NPS derived from a uniform patch. $PS_{Input}(u)$ and $PS_{Output}(u)$ are input and output test chart power spectra, and $u$ is spatial frequency.

$$MTF(u) = \sqrt{\frac{PS_{Output}(u) - NPS_{Output}(u)}{PS_{Input}(u)}} \quad (3)$$

The *intrinsic dead leaves implementation* for measuring the MTF [16] calculates signal transfer with respect to the cross spectrum. It is less susceptible to bias from noise since it measures the transfer of both image amplitude and phase content [13]. However, "reversible" image processes, such as sharpening or contrast stretching, are not accounted for [39]. These processes contribute heavily toward both the scene-dependent performance and the perceived image quality of non-linear systems.

Branca et al. derived MTFs from images of natural scenes [18] by adapting the direct dead leaves implementation (Eq. 3), so that $PS_{Input}(u)$ and $PS_{Output}(u)$ are the input and output scene power spectra, respectively. This method accounts for signal transfer scene-dependency but is often biased since: i) as with the direct dead leaves MTF, it compensates for the system's noise, $NPS_{Output}(u)$, using the NPS measured from a uniform patch. ii) input images are not windowed, or zero-padded, causing periodic replication artefacts that are discussed in the next section. The SPD-MTF measures presented in the next section address these limitations.

## Scene-and-Process-Dependent System Performance Measures

### *Scene-and-Process-Dependent NPSs (SPD-NPS)*

The SPD-NPS measurement framework, presented in Figure 3, derives the NPS using a number of captured images of the same scene, referred to as *replicates*. The 1D SPD-NPS is defined as the radial average of Eq. 2.1, where $\bar{g}(x, y)$ is the mean image of all replicates; other parameters are as previously stated.

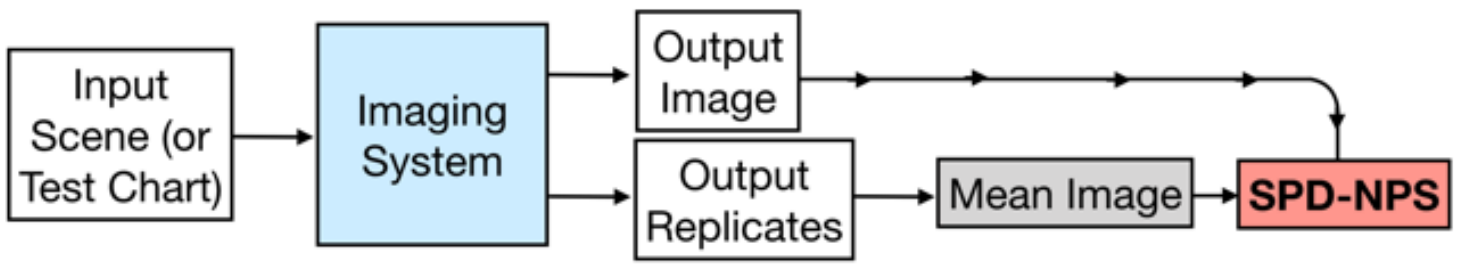

*Figure 3. The SPD-NPS measurement framework, adapted from* [3].

The framework accounts for system noise scene-dependency. It is computationally complex, however, since many replicates must be captured (10 in this paper); using fewer replicates underestimates system noise. The framework does not account for demosaicing artefacts of a fixed pattern, Figure 4, or sensor fixed pattern noise (FPN). The latter is less significant than temporally varying noise in current capture systems under most exposure conditions.

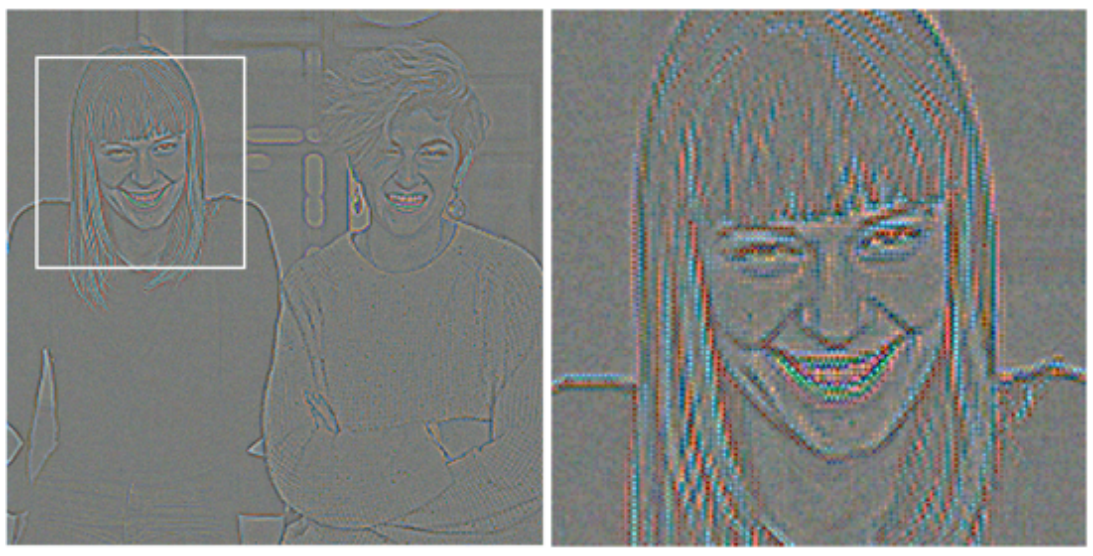

*Figure 4. Artefacts of fixed pattern caused by demosaicing the Branca et al.* [18] *'People' image (left), and detail of it (right); image contrast was enhanced.*

We define the four proposed SPD-NPS measures below.

i) The *dead leaves SPD-NPS* is derived from the dead leaves chart using the SPD-NPS framework (Figure 3). It aims to characterize average real-world system noise but uses assumptions of non-linear system behavior, discussed in the introduction.

ii) The *pictorial image SPD-NPS* is derived from a single pictorial scene using the SPD-NPS framework (Figure 3). It measures noise performance of the system with respect to that scene.

iii) The *mean pictorial image SPD-NPS* characterizes the average real-world noise performance of a system, accounting for its scene-dependency. It is defined as the mean of all pictorial image SPD-NPSs, $\bar{F}(s)$, over a set of $n$ images, as shown in Figure 1b. Averaging NPSs is unorthodox, but if the image set reflects the properties of commonly captured scenes, this measure tends toward a curve describing average real-world performance, as $n$ increases.

iv) The *pictorial image SPD-NPS standard deviation* describes the level of scene-dependent variation in a system's noise power. It is defined as the standard deviation of all pictorial image SPD-NPSs over a set of $n$ images. Its accuracy increases as *n* increases.

Measurement bias and variation error in the pictorial image SPD-NPSs are carried into the mean pictorial image SPD-NPS and pictorial image SPD-NPS standard deviation, respectively.

### *Scene-and-Process-Dependent MTFs (SPD-MTF)*

The SPD-MTF measurement framework, shown in Figure 5, builds upon the measurement method of Branca et al. [18], defined in the previous section. It accounts for system signal transfer as well as noise scene-dependency. It is defined using Eq. 3, where $PS_{Input}(u)$ and $PS_{Output}(u)$ are input and output scene (or dead leaves) power spectra, respectively, and $NPS_{Output}(u)$ is the pictorial image (or dead leaves) SPD-NPS.

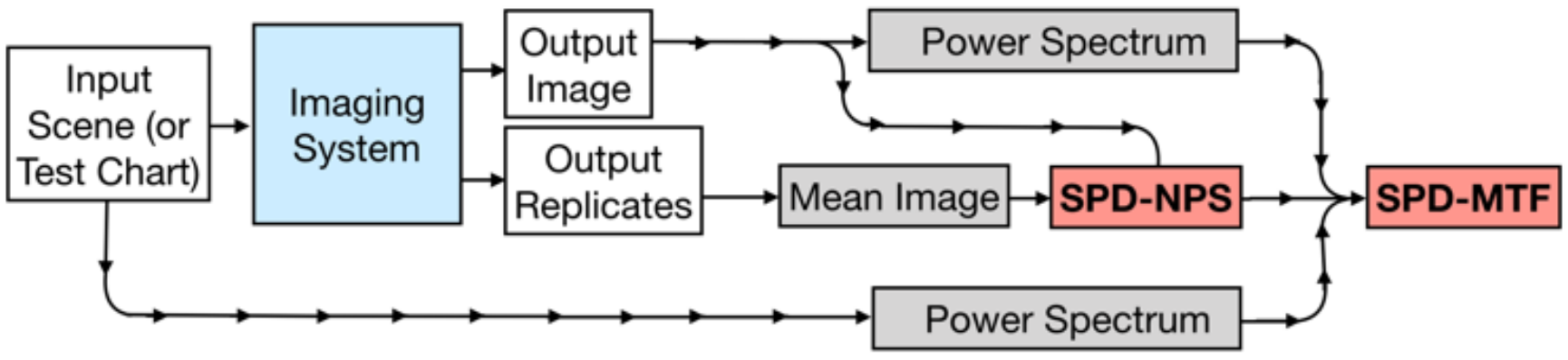

*Figure 5. The SPD-MTF measurement framework, adapted from* [3].

In the SPD-MTF measurements, input images are windowed to mitigate periodic replication artefacts, which are introduced to the 2D luminance spectrum during DFT processing when opposite image edges differ in luminance. Such artefacts not affected by image processing, and bias MTFs toward $MTF(u) = 1$ at all spatial frequencies. We applied a square-edged mask (Figure 6e), which tapers scene edges to a neutral pixel value using a cosine function of 1/128 cycles/pixel, starting at 64 pixels from each edge. The mask preserves scene signals where possible, to mitigate bias from signal-to-noise limitations, described below.

The SPD-MTF inherits signal-to-noise limitations from the direct dead leaves MTF that are described by Eq. 4. As the input image power, $PS_{Input}(u)$, approaches zero at frequency $u$, $MTF(u)$ becomes increasingly biased, especially if the measured NPS, $NPS_M(u)$, underestimates the real system NPS, $NPS_R(u)$. $PS_{Output}(u)$ is the output image power spectrum. Even if the real system NPS is measured with absolute accuracy in a theoretical ideal (i.e. $NPS_M(u)$ = $NPS_R(u)$) the numerator and denominator of line 3 of Eq. 4 limit toward zero at equal value, and $MTF(u)$ limits to 1. Thus, test images for the SPD-MTF should not have zero (or very low) power at any frequency. Power spectra should also be computed with a satisfactory precision.

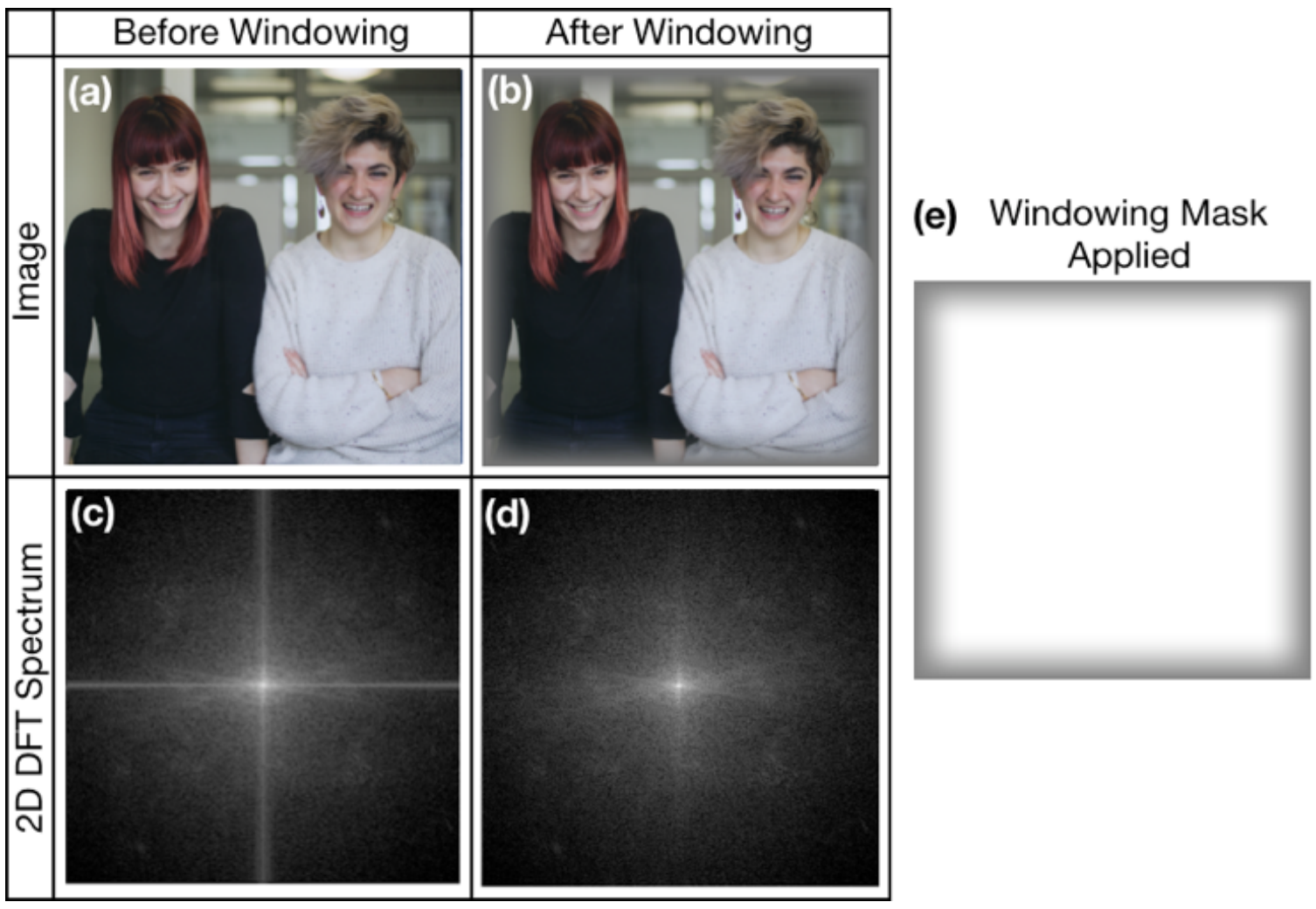

*Figure 6. 2D DFT log luminance spectra (c) and (d) for the Branca et al.* [18] *'People' image (a) and (b), before and after windowing, respectively, with (e).*

$$if \quad NPS_M(u) < NPS_R(u) \tag{4}$$

$$then \lim_{PS_{Input}(u)\to 0} PS_{Input}(u) < \ PS_{Output}(u) - NPS_M(u)$$

$$and \quad \lim_{PS_{Input}(u)\to 0} \left(\frac{PS_{Output}(u) - NPS_M(u)}{PS_{Input}(u)}\right) = \ \infty = \ MTF(u)$$

Signal-to-noise limitations and periodic replication artefacts explain the bias in the scene MTFs of Branca et al. [18], and why low-power scenes, and higher frequencies of generally lower power, were most affected. The SPD-MTF measurement framework is less biased since it implements a more appropriate noise measure, and windowing. Scenes with significant power across all frequencies, as well as the dead leaves chart, are generally expected to yield SPD-MTFs with acceptable levels of bias unless they are captured under very noisy exposure conditions. Low-power scenes are still expected to yield biased SPD-MTFs, particularly at higher frequencies, or if they are captured under noisy exposure conditions.

The four proposed SPD-MTF measures are defined below.

i) The *dead leaves SPD-MTF* characterizes the transfer of dead leaves signals by a system. It aims to describe average real-world signal transfer but requires assumptions of system behavior listed in the introduction. It applies the SPD-MTF framework (Figure 5) using dead leaves power spectra and the dead leaves SPD-NPS.

ii) The *pictorial image SPD-MTF* characterizes system signal transfer with respect to a given natural scene. It implements the SPD-MTF framework (Figure 5) using pictorial image power spectra and the pictorial image SPD-NPS.

iii) The *mean pictorial image SPD-MTF* characterizes the average real-world signal transfer of a system, accounting for system scene-dependency. It is defined as the mean of all pictorial image SPD-MTFs, over the same set of $n$ images used to compute the respective SPD-NPS measure. Bias is transferred to it from the pictorial image SPD-MTF.

iv) The *pictorial image SPD-MTF standard deviation* describes the level of scene-dependent variation in the signal transfer of a given system. It is defined as the standard deviation of all pictorial image SPD-MTFs over the set of $n$ images. Variation error in the pictorial image SPD-MTFs causes it to be biased.

## Camera System Simulation and Test Images

Two image capture system pipelines were simulated in MATLAB$^{TM}$ and tuned to behave like real camera-phone cameras, under various simulated exposure conditions. Both pipelines modelled physical capture processes and image pre-processing identically, in the following order.

Lens blur was simulated by convolution with a Gaussian approximation for the central lobe of a diffraction-limited lens airy disk, using the $f$-number and the pixel pitch of an *iPhone 6* smartphone [40]. 2D photon noise was simulated at linear signal-to-noise ratios (SNR) of 5, 10, 20 and 40 at saturation, using Poisson statistics [41]. It was scaled by factors of 2, 1, and 3.3 in the R, G, and B channels, respectively, to account for the lower quantum efficiency of the R and B channels. Dark current noise was modelled as Gaussian noise with a higher standard deviation at lower SNRs. The noise floor was then removed, and highlights were recovered by black and white level adjustments, respectively. Pixel information was sampled according to a 'grbg' color filter array. Most simulations implement the latter process before the simulation of noise and pre-processing. But the chosen order produced the same output images to this order and facilitated the scaling of Photon noise in the R and B channels.

Further, the non-linear pipeline implemented the following non-linear content-aware image signal processing algorithms, in order. Demosaicing was by the One Step Alternating Projections (OSAP) [42] algorithm, set to 'full convergence'. Denoising was by Block Matching and 3D Filtering (BM3D) [24] using the 'normal' profile. Images from each color channel were then sharpened by the Guided Image

Filter (GIF) [36] and concatenated. The latter process used the input image to the filter as the guidance image [36].

The linear pipeline applied the following equivalent linear algorithms, in order. Demosaicing was by the Malvar et al. [43] algorithm. Denoising was by 2D spatial domain Gaussian filtering. Sharpening was by the *imsharpen* MATLAB$^{TM}$ function.

Output images from both pipelines were then saved in a lossless Portable Network Graphics (PNG) [44] file format.

The image set consisted of 50 high-quality imaged scenes of 512 by 512 pixels, from the LIVE Image Quality Assessment Database [45], Branca et al. [18], Fry et al. [46] and Allen et al. [47] image sets. They covered a wide range of subjects and reflected the variety of scenes captured by contemporary camera systems.

## Validation and Discussion

The SPD-MTF and SPD-NPS measures were validated by analyzing measurements from both the linear and non-linear pipelines, at all photon noise SNRs; here, we present results from SNR levels of 40 and 5. The chosen SNRs represent very good and very poor signal quality, respectively. SNRs in between showed comparable trends. The actual SNRs of the processed images are higher due to dark current noise. All NPSs describe luminance noise. Burns' direct dead leaves MTF implementation [48] was adapted for computing all measures. All measurements were smoothed by a moving average filter of seven segments.

### *Scene-and-Process-Dependent NPS (SPD-NPS)*

Figures 7 and 8 show SPD-NPSs derived from the dead leaves chart and pictorial images, respectively. Although there is no current way of deriving the ground truth (or "correct") NPS for a given system, Figure 7 demonstrates that the dead leaves SPD-NPS (red lines) characterizes the average real-world noise performance of the non-linear pipeline more competently than the uniform patch NPS (black line), validating hypothesis 1. Levels of bias in both measures are similar. This is because their respective measurements from the linear pipeline are alike, as would be expected, in theory, for two NPS measures reliant on linear system theory (see Figure 1a). The uniform patch NPS is derived from a less suitable input signal than the dead leaves SPD-NPS and should be considered the less representative, thus less ”correct” measure. It underestimates the latter after non-linear denoising, as shown in Figures 7d and 7j. This underestimation is compounded slightly by sharpening. It is also greater at higher SNRs because the less intensive denoising still cleans the uniform patch effectively. Lowering the number of replicates from 100 to 10 does not alter the dead leaves SPD-NPS significantly. Measurements derived using 10 and 100 replicates are difficult to distinguish from one another on logarithmic axes. SPD-NPSs derived from pictorial images (Figure 8) were thus computed with ten replicates to maintain consistency.

Figure 8 indicates that measurement bias is similar for the pictorial image SPD-NPSs (grey lines), mean pictorial image SPD-NPS (black line), and dead leaves SPD-NPS (red line), since their respective measurements from the linear pipeline are alike.

The pictorial image SPD-NPS (Figure 8, grey lines) is validated as the most suitable measure for non-linear system noise with respect to a given input scene (hypothesis 3). It accounts for scene-dependent system behavior, as demonstrated by the variation in measurements after non-linear denoising (see Figures 8d and 8j).

The average real-world noise performance of the pipelines is best described by the mean pictorial image SPD-NPS (Figure 8, black line), validating hypothesis 4. This measure accounts for the general trends in scene-dependent system behavior over the various test scenes. The dead leaves SPD-NPS (Figure 8, red line) generally underestimates the noise power measurements derived from real scenes, suggesting non-linear denoising is capable of removing noise in dead leaves signals more effectively than in pictorial scenes.

The pictorial image SPD-NPS standard deviation (Figure 8, black dotted lines) describes well the general level of scene-dependency in the noise of both pipelines in real-world capture scenarios (hypothesis 5). Non-linear denoising is the primary cause of scene-dependency. Non-linear sharpening does not compound the spread of the pictorial image SPD-NPSs (grey lines) but does change their shape and order in a scene-dependent manner. The pictorial image SPD-NPS standard deviation accounts for the former, but not the latter. Note that, the sudden change in the lower standard deviation boundary in Figure 8j is not a discontinuity. It is caused by the curve crossing the x-axis of a graph with a logarithmic y-axis.

We have noted that scene-dependent variation is also visible in the pictorial image SPD-NPSs from the linear pipeline, when Figure 8 is plotted on linear y-axes, and have analyzed this variation further in reference [49]. Its standard deviation is around 10% and 15% of the mean pictorial image SPD-NPS at SNRs 40 and 5, respectively. It is assumed to have been caused by the scene-dependent variations in Poisson noise, the scaling of noise to simulate color channel quantum efficiency, and the black/white level adjustments. It also may be due to variations in the level of measurement error, which cannot be distinguished straightforwardly from any curve variations caused by the above genuine sources of system scene-dependency. Regardless of the origin of this variation, our simulations suggest it should not significantly affect the validity of the pictorial image SPD-NPS, and the related mean pictorial image SPD-NPS and pictorial image SPD-NPS standard deviation measures.

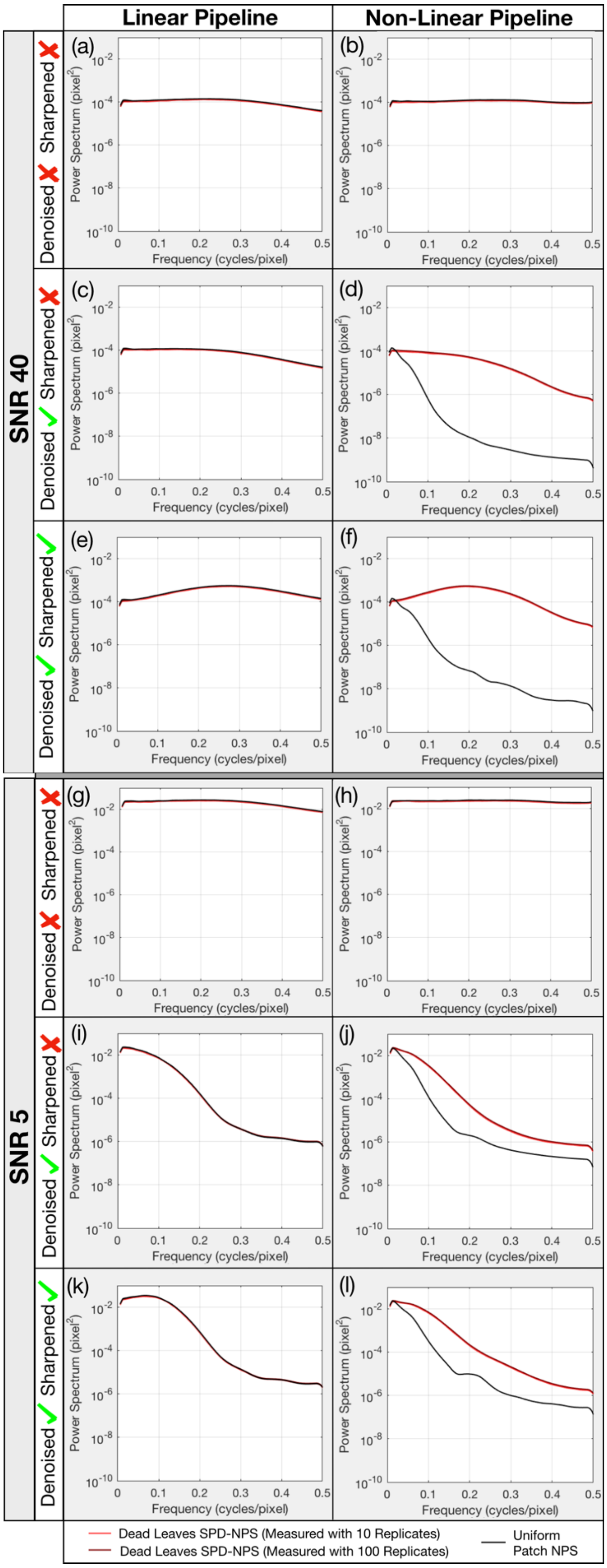

*Figure 7. Luminance NPSs derived from the dead leaves chart (red lines) and uniform patches (black lines) at different stages of processing at SNR 40 (a) – (f) and SNR 5 (g) – (l).*

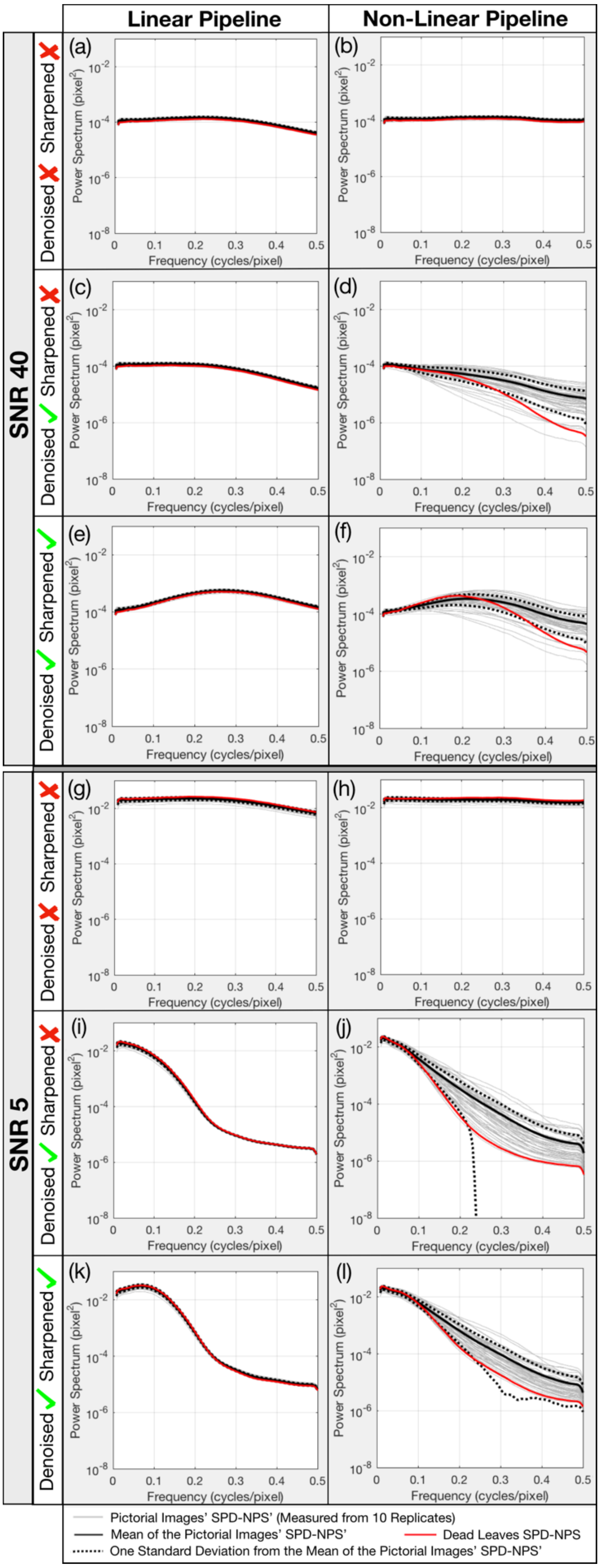

*Figure 8. Pictorial image SPD-NPSs (grey lines), mean pictorial image SPD-NPSs (black lines), SPD-NPS standard deviations (black dotted lines), and dead leaves SPD-NPSs (red lines) of luminance noise at different stages of processing at SNRs 40 and 5.*

### *Scene-and-Process-Dependent MTF (SPD-MTF)*

Similarly to the NPS measurement, there is no current method to derive the ground truth, or "correct", MTF of a system. Thus, we choose in this paper to validate the various SPD-MTFs by comparing their behavior with the direct dead leaves MTF [15]. Figures 9 and 10 support the validation of SPD-MTF measures from the dead leaves chart and pictorial images, respectively. All measurements are heavily biased before denoising at SNR 5, due to signal-to-noise limitations, as described by Eq. 4. This is because noise power is extremely high at these SNR levels and is thus underestimated significantly by all NPS measures. For the SPD-NPS measures, increasing the number of replicates reduced this bias, as shown in Figure 9. Denoising also mitigated bias in all noise measures and is generally applied in such situations. Thus, Figures 9g and 9h and Figures 10g and 10h are referred to from here on as less relevant conditions.

The dead leaves SPD-MTF (Figure 9, red lines) employs a more appropriate noise measure for non-linear systems than the direct dead leaves MTF (Figure 9, black line) as confirmed by observations from Figure 7. This causes it to characterize signal transfer in the non-linear pipeline with slightly improved accuracy, under the most relevant conditions (hypothesis 2). Both the above measures have similar levels of bias since measurements from the linear pipeline are alike under such conditions. Reducing the number of replicates from 100 to 10 increases the underestimation of noise by the dead leaves SPD-NPS, introducing minor positive bias to dead leaves SPD-MTFs at higher frequencies due to signal-to-noise limitations (Eq. 4). Ten replicates were considered sufficient for the calculation of the SPD-MTFs shown in Figure 10.

Observations from Figure 10 strongly suggest the pictorial image SPD-MTF (grey lines) is the most appropriate measure for non-linear system signal transfer concerning a given input scene (hypothesis 3). Scene-dependent variation in this measure is significantly higher after denoising and sharpening by non-linear algorithms than by the equivalent linear algorithms. This demonstrates that pictorial image SPD-MTFs account for relevant scene-dependent system behavior.

However, scene-dependent variation is also significant in the pictorial image SPD-MTF measurements from the linear pipeline, under some conditions. This is partly due to genuine scene-dependency resulting from black/white level adjustments and scaling of noise to model color channel quantum efficiency variations. However, we expect this variation to be mainly caused by bias from signal-to-noise limitations. This bias is scene-dependent, due to variations in image signal power, as well as variations in any underestimation of noise power by the employed SPD-NPS measures. Higher frequencies are generally most biased, especially in low-power images, at lower SNRs. This bias is mitigated by denoising, although non-linear denoising introduces significant scene-dependencies of its own. It is not presently possible to distinguish between the scene-dependent variation in the pictorial image SPD-MTFs that results from this bias, and the variation that results from genuine system scene-dependency (i.e. interaction between the signal of the scene and the employed image processing algorithms).

When it is necessary to describe the signal transfer of a non-linear system with respect to a given scene – for example, in image quality modelling applications – the pictorial image SPD-MTF is, in theory, the most valid MTF measure since it uses an appropriate input signal and accounts most comprehensively for system scene-dependency. However, in practice, deriving the MTF from pictorial input signals often increases measurement error, which mainly results from a lack of image power at a given frequency triggering signal-to-noise limitations in the calculation. This affects the "correctness" of the measure and should be investigated further.

The pictorial image SPD-MTF standard deviation (Figure 10, black dotted lines) describes the overall level of system scene-dependency effectively (hypothesis 5). The influence of the above scene-dependent bias in the pictorial image SPD-MTF causes this scene-dependency measure to be larger for the linear system than would be expected in theory, and as demonstrated in practice by the equivalent SPD-NPS measure (Figure 8, black dotted lines). Nevertheless, we expect that it is a valuable measure, especially if bias in the pictorial image SPD-MTF can be reduced further. Like the equivalent SPD-NPS measure, it accounts for the spread of the pictorial image SPD-MTF curves, but not changes in their shape, or order.

The mean pictorial image SPD-MTF (Figure 10, black line) is validated as more relevant than the dead leaves SPD-MTF (Figure 10, red line) for the measurement of the average real-world signal transfer of non-linear systems (hypothesis 4). Interestingly, the levels of bias are similar for both measures under the most relevant conditions. This is because, for the linear pipeline, calculating the mean of the pictorial images' SPD-MTFs averages out their scene-dependent variation and bias, yielding a curve of similar shape to the dead leaves SPD-MTF. This would suggest that – providing it is measured from an image set consisting of commonly captured scenes – the mean pictorial image SPD-MTF is not affected significantly by the bias in the individual pictorial image SPD-MTF measurements it is derived from.

After non-linear denoising and/or sharpening, however, the dead leaves SPD-MTF is dissimilar to the various pictorial image SPD-MTFs, and often underestimates the mean pictorial image SPD-MTF; both these other measures account for scene-dependency more suitably than the dead leaves SPD-MTF. We infer that the non-linear content-aware denoising and sharpening algorithms process dead leaves signals unlike signals from the average pictorial scene. For fairness of comparison, the dead leaves chart was windowed as per all scenes, to mitigate bias from periodic replication artefacts. This explains the minor differences between the respective dead leaves SPD-MTFs in Figures 9 and 10.

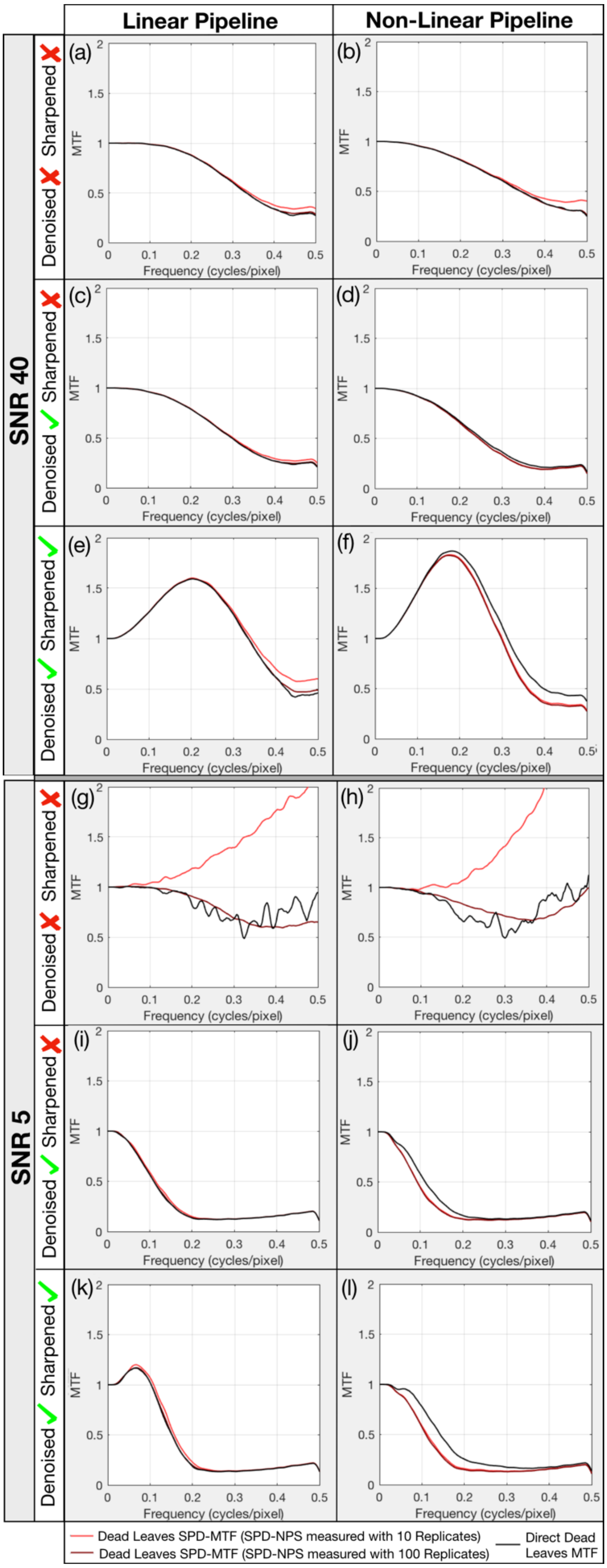

*Figure 9. Dead leaves SPD-MTFs (red lines) and direct dead leaves MTFs (black lines) and at different stages of processing, at SNR 40 (a) – (f) and SNR 5 (g) – (l). Windowing was not applied for these measurements.*

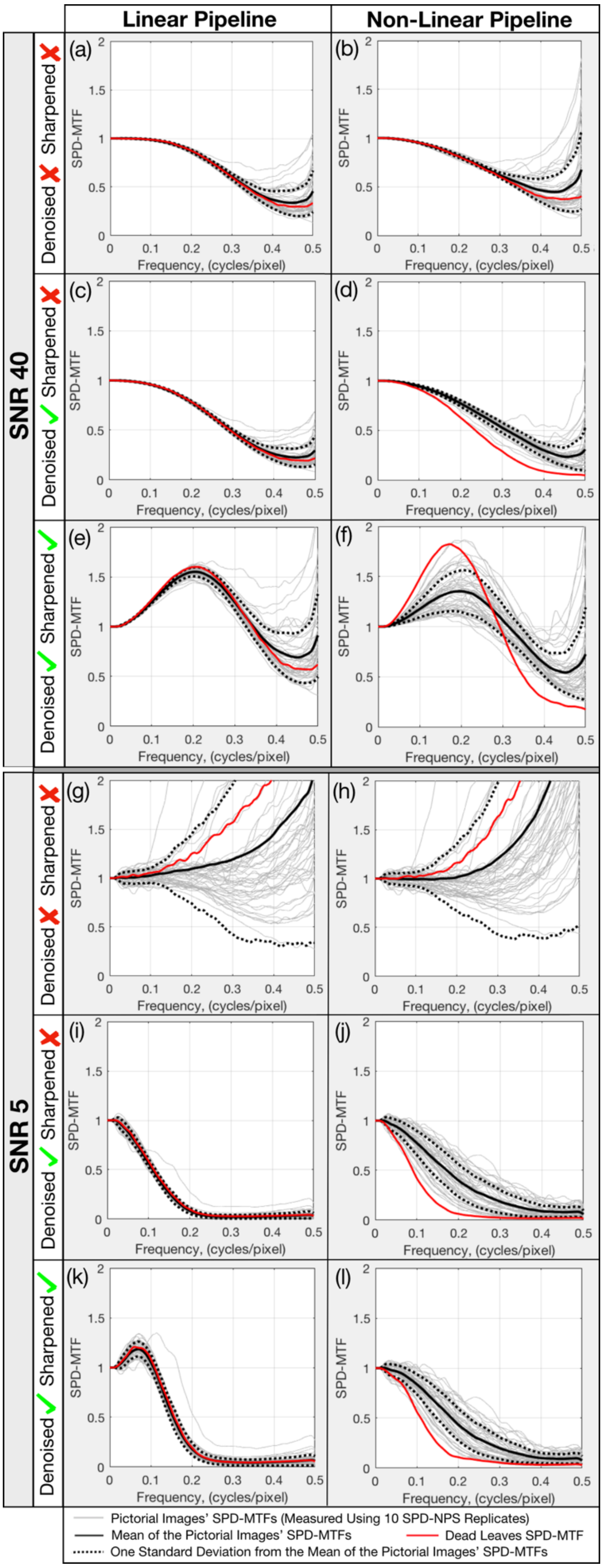

*Figure 10. Pictorial image SPD-MTFs (grey lines), mean pictorial image SPD-MTFs (black lines), SPD-MTF standard deviations (black dotted lines) and dead leaves SPD-MTFs (red lines), at different stages of processing at SNRs 40 and 5. All scenes and test charts were windowed for these measurements.*

## Conclusions

In this paper, we have reviewed the limitations of current MTF and NPS measures when they are used to characterize capture systems that incorporate non-linear, content-aware spatial image signal processes. We further introduced a number of scene-and-process-dependent MTF (SPD-MTF) and NPS (SPD-NPS) measures, developed to account for the scene-dependency of such systems. These measures were validated using simulated camera-phone image capture pipelines. The fact that they have revealed significant scene-dependency in the performance of the non-linear pipeline indicates their promise for the characterization of real camera systems and/or other non-linear imaging systems.

The *pictorial image SPD-MTF* and *SPD-NPS* measured pipeline performance suitably with respect to a given input scene, accounting for scene-dependent behavior. They are the only current measures capable of such characterization. In a separate publication [19], we demonstrate that these measures are suitable input parameters for engineering IQMs, for applications related to the modelling of the perceived image quality of a given input scene.

The *mean pictorial image SPD-MTF* and *SPD-NPS* characterized the average real-world performance of the pipelines, accounting for general trends in pipeline scene-dependency. We propose that these measures are used as performance optimization parameters, and input parameters for IQMs, for applications concerning the average real-world image quality of systems. Current equivalent measures do not account for system scene-dependency.

The *pictorial image SPD-MTF* and *SPD-NPS standard deviation* described appropriately the level of scene-dependent variation in the performance of each pipeline. They are the only current measures that can describe system scene-dependency, but do not account for all aspects of it. In a parallel paper [49], we examine whether the latter affects their validity. We propose they should be employed in conjunction with the mean pictorial image SPD-MTF and SPD-NPS, thus characterizing both the average real-world performance of the system, as well as its level of scene-dependency.

The *dead leaves SPD-MTF* and *SPD-NPS* measured non-linear pipeline performance more accurately than the current direct dead leaves MTF and uniform patch NPS, respectively. We propose these measures should be used to estimate the average real-world system performance. However, results from these measures were generally outliers, when compared to the measures derived from pictorial images. This is because non-linear content-aware image signal processing algorithms were triggered at different levels by the dead leaves chart, compared to natural scene signals. The dead leaves chart simulates an average scene signal with a typical power spectrum. However, with respect to these algorithms, it is only a mathematically generated image with limited relation to the complex spatial signals in pictorial scenes. We propose a more suitable test chart should either simulate natural scene signal structure more comprehensively or be composed of such signals.

The SPD-NPS measures showed little measurement error. Thus, they are more appropriate than current equivalent measures (if equivalent measures exist) for non-linear content-aware capture systems, since they account for the effect of relevant input signals on the noise power of the system. However, they do not account for fixed pattern noise/artefacts. Their requirement for many replicates also causes all SPD-MTFs and SPD-NPSs to be more computationally complex than current measures.

The SPD-MTF measures are also more suitable for non-linear content-aware capture systems than current equivalent measures. However, the pictorial image SPD-MTF was prone to measurement bias under certain conditions, due to signal-to-noise limitations inherited from the direct dead leaves MTF implementation. This bias was mitigated by denoising, or by computing the measure using more replicates, or with scenes of higher signal power. Further investigations are recommended to reduce it further.

This bias was scene-dependent and cannot currently be distinguished from genuine system scene-dependency that results from interactions between the image signal and non-linear image processing. The fact that it was scene-dependent meant that this bias affected the accuracy of the pictorial image SPD-MTF standard deviation. However, the effect of this bias averaged out over the 50 test scenes to comparable levels to bias in MTF measurements derived from dead leaves signals. Consequently, we expect the accuracy of the mean pictorial image SPD-MTF was less affected by this bias than the other SPD-MTF measures derived from pictorial scenes.

In-depth analysis of this bias, as well as other measurement errors in the SPD-MTF and SPD-NPS, were beyond the scope of this paper. We suggest that further analysis of such error is needed to evaluate the "correctness" of each SPD-MTF and SPD-NPS measure. In the case of the SPD-MTFs, this may involve adapting error propagation methods for existing dead leaves MTF measurement implementations [50].

All the proposed SPD-MTF and SPD-NPS measures would benefit from validation with real capturing systems. In a separate paper [19], we have used the various measures as input parameters to engineering IQMs, with the aim of modelling the quality of individual scenes – the modified IQMs are generally more successful (i.e. they correlate more accurately with the perceived quality of each output scene). We have also developed novel and competitive spatial IQMs using these various measures. In a further parallel publication [49], we have evaluated in detail and quantified the level of scene-dependency in linear and non-linear camera pipelines using the various measures, and have validated relevant single-figure metrics for objective system performance.

## Author Biography

Edward Fry is a PhD. student at the University of Westminster, UK, in the field of imaging systems performance measurement and image quality modeling. His research focuses on scene-dependency in imaging systems performance and its impact upon overall perceived image quality.